\documentclass{article} 
\usepackage{amsmath,amssymb,amsfonts}   
\usepackage{graphicx,bm}

\usepackage{amsmath,amssymb,amsfonts}   
\usepackage{graphicx,bm}

 \usepackage{tikz-cd}

\newcommand{\calU}{{\mathcal U}}

\newcommand{\calP}{{\mathcal P}}
\newcommand{\calQ}{{\mathcal Q}}

\newcommand{\calF}{{\mathcal F}}

\newcommand{\calM}{{\mathcal M}}

\newcommand{\R}{{\mathbb R}}
\renewcommand{\L}{{\mathbb L}}
\newcommand{\X}{\mathbf{X}}
\renewcommand{\P}{\mathbb{P}}
\newcommand{\PP}{\widetilde{P}}
\newcommand{\Q}{\widetilde{Q}}
\newcommand{\x}{\mathbf{x}}

\newcommand{\e}{{\mathrm e}}

\newcommand{\n}{\mathbf n}
\newcommand{\calT}{{\mathcal T}}
\renewcommand{\P}{\mathbb P}
\newcommand{\p}{\widetilde{p}}

\newcommand{\q}{\widetilde{q}}

\renewcommand{\S}{\widetilde{S}}
\newcommand{\ellh}{\widehat{\ell}}

\begin{document}

 \title{Spectral theory of diffusion in partially absorbing media}

\author{ \em
P. C. Bressloff, \\ Department of Mathematics, 
University of Utah \\155 South 1400 East, Salt Lake City, UT 84112}

\maketitle

\begin{abstract} 
A probabilistic framework for studying single-particle diffusion in partially absorbing media has recently been developed in terms of an encounter-based approach. The latter computes the joint probability density (generalized propagator) for particle position $\X_t$ and a Brownian functional ${\mathcal U}_t$ that specifies the amount of time the particle is in contact with a reactive component $\calM$. Absorption occurs as soon as $\calU_t$ crosses a randomly distributed threshold (stopping time). Laplace transforming the propagator with respect to $\calU_t$ leads to a classical boundary value problem (BVP) in which the reactive component has a constant rate of absorption $z$, where $z$ is the corresponding Laplace variable. Hence, a crucial step in the encounter-based approach is finding the inverse Laplace transform. In the case of a reactive boundary $\partial \calM$, this can be achieved by solving a classical Robin BVP in terms of the spectral decomposition of a Dirichlet-to-Neumann operator. In this paper we develop the analogous construction in the case of a reactive substrate $\calM$. In particular, we show that the Laplace transformed propagator can be computed in terms of the spectral decomposition of a pair of Dirichlet-to-Neumann operators. However, inverting the Laplace transform with respect to $z$ is more involved. We illustrate the theory by considering a 1D example where the Dirichlet-to-Neumann operators reduce to scalars.

\end{abstract}

\section{Introduction}
\begin{figure}[t!]
\centering
  \includegraphics[width=11cm]{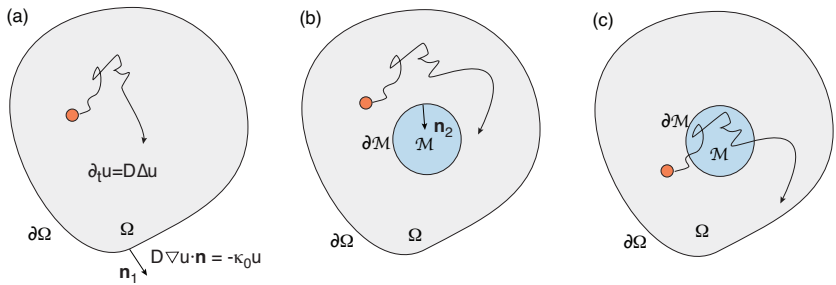}
  \caption{(a) Diffusion of a particle in a bounded domain $\Omega$ with a partially absorbing boundary $\partial \Omega$. The probability of particle absorption depends on the amount of time spent in a neighborhood of $\partial \Omega$, which is specified by the local accumulation time $\ell_t$. (b) Diffusion in the bounded domain $\Omega\backslash  \calM$, with $\partial \calM$ acting as a partially absorbing boundary. (c) A particle diffusing in $\Omega $ can freely enter and exit the substrate domain $\calM$. The probability of particle absorption depends on the amount of time spent within $\calM$, which is specified by the occupation time $A_t$. Note that the unit normal $\n_2$ is directed towards the interior of $\calM$ in (b).}
  \label{fig1}
\end{figure}

The three classical boundary conditions for the diffusion equation $\partial u /\partial t=D\nabla^2 u $ in a bounded domain $\Omega$ are, respectively, Dirichlet ($u(\x,t)=0$), Neumann ($\nabla u(\x,t) \cdot \n_1=0$) and Robin ($D\nabla u(\x,t)\cdot \n_1+\kappa_0 u(\x,t)=0$) for all $\x\in \partial \Omega$. Here $u$ is particle concentration, $D$ is the diffusivity, $\kappa_0$ is a positive reactivity constant, and $\n_1$ is the outward unit normal at a point on the boundary, see
Fig. \ref{fig1}(a). However, implementing these boundary conditions at the level of a single diffusing particle is non-trivial. Individual trajectories of the particle are generated by a stochastic differential equation (SDE) that, in the case of pure diffusion in $\R^d$ is given by a Wiener process. Although the evolution of the probability density $p(\x,t)$ for the random position $\X_t $ in a bounded domain $\Omega$ is identical to the macroscopic diffusion equation for $u$, the effect of the boundary on the underlying SDE is more complicated. The simplest case is a totally absorbing boundary (Dirichlet), which can be handled by stopping the Brownian motion on the first encounter between particle and boundary. The random time at which this event occurs is known as the first passage time (FPT). On the other hand, it is necessary to modify the stochastic process itself in the case of a totally or partially reflecting boundary. For example, one can implement a Neumann boundary condition by introducing a Brownian functional known as the boundary local time \cite{Levy39,McKean75,Freidlin85,Majumdar05}. The latter determines the amount of time that a Brownian particle spends in the neighborhood of points on the boundary. Probabilistic versions of the Robin boundary condition can also be constructed \cite{Papanicolaou90,Milshtein95,Singer08}.

One of the assumptions of the Robin boundary condition is that the surface reactivity is a constant. However, various surface-based reactions are better modeled in terms of a reactivity that is a function of the local time \cite{Bartholomew01,Filoche08}. 
That is, the surface may need to be progressively
activated by repeated encounters with a diffusing
particle, or an initially highly reactive surface may become less active due to multiple interactions with the particle (passivation). Recently, a theoretical framework for analyzing a more general class of partially absorbing boundary has been developed using a so-called encounter-based approach \cite{Grebenkov19b,Grebenkov20,Grebenkov22}. The basic idea is to consider the joint probability density or generalized propagator $P(\x,\ell,t)$ for the pair $(\X_t,\ell_t)$ in the case of a perfectly reflecting boundary, where $\X_t$ and $\ell_t$ denote the particle position and local time, respectively. The effects of surface reactions are then incorporated 
 by introducing the stopping time 
$
{\mathcal T}=\inf\{t>0:\ \ell_t >\widehat{\ell}\}$,
 with $\widehat{\ell}$ a so-called stopping local time. Given the probability distribution $\Psi(\ell) = \P[\ellh>\ell]$, the marginal probability density for particle position is defined according to
 $  p(\x,t)=\int_0^{\infty} \Psi(\ell)P(\x,\ell,t)d\ell$. The classical Robin boundary condition for the diffusion equation corresponds to the exponential distribution
$\Psi(\ell) =\e^{-\gamma \ell}$, where $\gamma =\kappa_0/D$. On the other hand, if $\kappa=\kappa(\ell)$ then $\Psi(\ell)=\exp\left (-D^{-1}\int_0^{\ell}\kappa(\ell')d\ell'\right )$. The crucial step in the encounter-based approach is computing the generalized propagator
$P(\x,\ell,t)$ by solving a corresponding boundary value problem (BVP) \cite{Grebenkov20,Grebenkov22}. Performing a double Laplace transform with respect to the time $t$ and the local time $\ell$,
 \begin{equation}
 \label{dLT}
 \calP(\x,z,s)\equiv \int_0^{\infty}\e^{-z\ell}\int_0^{\infty}\e^{-st}P(\x,\ell,t)dtd\ell,
 \end{equation}
one finds that  $\calP(\x,z,s)$ satisfies a modified Helmholtz equation with a Robin boundary condition on $\partial \Omega$, in which the effective reactivity is proportional to the Laplace variable $z$. Hence, the calculation of the Laplace transformed propagator $\PP(\x,\ell,s)$ reduces to solving a classical Robin BVP and then inverting the solution with respect to the Laplace variable $z$,
\begin{equation}
 \label{sLT}
 \PP(\x,\ell,s)\equiv \int_0^{\infty}\e^{-st}P(\x,\ell,t)dt={\mathcal L}_{\ell}^{-1}[\calP(\x,z,s) ] .\end{equation}
It turns out that solving the Robin BVP in terms of the spectrum of an associated Dirichlet-to-Neumann operator yields a series expansion that is easily inverted with respect to the Laplace variable $z$ \cite{Grebenkov20}. The corresponding marginal density $\p(\x,s)=\int_0^{\infty} \Psi(\ell) \PP(\x,\ell,s)d\ell$ and the associated boundary flux generate various quantities of interest without having to transform back to the time domain, including the mean first passage time (MFPT) for absorption.

A simple generalization of diffusion in a bounded domain $\Omega$ is illustrated in Fig. \ref{fig1}(b), whereby a target domain $\calM$ is inserted within the interior of $\Omega$. There is now an exterior boundary $\partial \Omega$ and an interior boundary $\partial \calM$, each of which has an associated boundary condition. A standard scenario at the single particle level is taking $\partial \Omega$ to be totally reflecting, or setting $\Omega=\R^d$, and calculating the statistics of the first absorption time in the case of a totally or partially absorbing target boundary $\partial \calM$. However, the possibility of partial absorption naturally leads to another generalization, as illustrated in Fig. \ref{fig1}(c), in which $\calM$ acts as a partially absorbing substrate. Now the particle can freely enter and exit the domain $\calM$, and the probability of being absorbed depends on the amount of time spent within $\calM$. The latter is specified by another Brownian functional known as the occupation time $A_t$ \cite{Majumdar05}. We have recently shown how to extend the encounter-based approach of Grebenkov to partially absorbing substrates by constructing the generalized propagator for the occupation time $A_t$ rather than the local accumulation time $\ell_t$ \cite{Bressloff22a}. In particular, we used a Feynman-Kac formula to derive the BVP for the occupation time propagator and solved the resulting BVP in the special case of a spherically symmetric domain. However, more general aspects of the theory were not explored.

In this paper we develop the general analysis of the occupation time propagator BVP. In section 2 we describe the encounter-based method for analyzing single-particle diffusion in partially absorbing media \cite{Grebenkov20,Grebenkov22,Bressloff22a}. We follow the particular formulation developed in Ref. \cite{Bressloff22a}, which focuses on the construction of the propagator BVPs for the local time on $\partial \calM$ and the occupation time within $\calM$, respectively. We show how to incorporate partial absorption in terms of a stopping time condition, and derive a general formula for the mean first passage time (MFPT) in terms of the corresponding survival probability. In section 3 we perform a double Laplace transform with respect to $t$ and the occupation time $A_t$, and derive the analog of Robin boundary conditions for the occupation time BVP, namely, the particle can be absorbed at a constant rate $z$ within $\calM$, where $z$ is the Laplace variable conjugate to the occupation time. In section 4 we explore to what extent spectral methods used to solve the Robin BVP for the local time propagator \cite{Grebenkov20} can be carried over to the occupation time propagator BVP. We establish that the solution can be computed in terms of the spectral decomposition of a pair of Dirichlet-to-Neumann operators. However, inverting the Laplace transform in $z$ is more complicated. Finally, we illustrate the theory by considering a partially absorbing substrate in 1D for which the Dirichlet-to-Neumann operators reduce to scalars. We derive an explicit expression for the propagator and use this to determine the MFPT for absorption.

\setcounter{equation}{0}

  \section{Brownian functionals and the generalized propagator BVP}
  
Consider a particle diffusing in a bounded domain $\Omega\subset \R^d$ with a totally reflecting boundary $\partial \Omega$.  
 Let $\X_t$ denote the position of the particle at time $t$.
A Brownian functional over a fixed time interval $[0,T]$ is defined as a random variable $\calU_t$ given by \cite{Majumdar05}
\begin{equation}
\calU_t=\int_0^t F(\X_{\tau})d\tau,
\end{equation}
where $F(\x)$ is some prescribed function or distribution such that $\calU_t$ has positive support and $\X_0=\x_0$ is fixed. We also assume that $\calU_0=0$. Let $P(\x,u,t|\x_0)$ denote the joint probability density or propagator for the pair $(\X_t,\calU_t)$. It follows that
 \begin{eqnarray}
 \label{A1}
& P(\x,u,t|\x_0)=\bigg \langle \delta\left (u -\calU_t \right )\bigg \rangle_{\X_0=\x_0}^{\X_t=\x} ,
 \end{eqnarray}
 where expectation is taken with respect to all random paths realized by $\X_{\tau}$ between $\X_0=\x_0$ and $\X_t=\x$. 
 Using the Feynman-Kac formula, it can be shown that the propagator satisfies a BVP of the form \cite{Bressloff22a}
\begin{subequations}
\label{calP}
\begin{align}
\frac{\partial P(\x,u,t|\x_0)}{\partial t}&=D\nabla^2 P(\x,u,t|\x_0)-F(\x) \frac{\partial P}{\partial u}(\x,u,t|\x_0) \nonumber \\ &\quad - \delta(u)F(\x)P(\x,0,t|\x_0),\  \x\in \Omega,  \\
&\nabla P(\x,u,t|\x_0)\cdot \n_1 =0,\ \x \in \partial \Omega. 
\end{align}
\end{subequations}
Here $\n_1$ is the outward unit normal to a point on the boundary $\partial \Omega$. If $\Omega$ is unbounded then we replace the Neumann condition (\ref{calP}b) by $P(\x,u,t|\x_0)\rightarrow 0$ as $|\x|\rightarrow \infty$. The initial condition is $P(\x,u,0|\x_0)=\delta(\x-\x_0)\delta(u)$.

Suppose that a target domain $\calM$ is introduced within the interior of $\Omega$ along the lines of Fig. \ref{fig1}. There are then two important Brownian functionals that can be associated with $\calM$ \cite{Majumdar05}. The first is the boundary local time $\ell_t$, which applies when the interior boundary $\partial \calM$ is totally reflecting:
\begin{equation}
\label{loc}
\ell_t=\lim_{h\rightarrow 0} \frac{D}{h} \int_0^tH(h-\mbox{dist}(\X_{\tau},\partial \calM))d\tau,
\end{equation}
where $H$ is the Heaviside function. Note that although $\ell_t$ has units of length due to the additional factor of $D$, it essentially specifies the amount of time that the particle spends in an infinitesimal neighborhood of the surface $\partial \Omega$. It is clear from definition (\ref{loc}) that $\ell_t$ is a non-decreasing stochastic process, which remains at zero until the first encounter with the boundary. One well known property of reflected Brownian motion is that when a particle hits a smooth surface, it returns to the surface an infinite number of times within an infinitely short time interval. Although each of these returns generates an infinitesimal increase in the boundary local time, the net effect of multiple returns is a measurable change in $\ell_t$. In a real physical system, there is a natural surface boundary layer of width $\delta$ that is determined by short-range atomic interactions. One can then approximate the boundary local time by the residence time of the particle in the boundary layer. An analogous regularization occurs in numerical simulations due to spatial discretization. As highlighted in Ref. \cite{Grebenkov20}, the boundary local time is a more universal quantity since it is independent of the boundary layer width, and is thus easier to deal with mathematically. The second Brownian functional is the occupation time $A_t$, which applies when the particle can freely enter and exit $\calM$:
\begin{equation}
\label{occ}
A_t=\int_{0}^tI_{\calM}(\X_{\tau})d\tau .
\end{equation}
Here $I_{\calM}(\x)$ denotes the indicator function of the set $\calM\subset \Omega$, that is, $I_{\calM}(\x)=1$ if $\x\in \calM$ and is zero otherwise. Clearly $A_t$ specifies the amount of time the particle spends within $\calM$ over the interval $[0,t]$.

We now consider the propagator BVP for each of the above two Brownian functionals. In the case of the local time (\ref{loc}), the effective bounded domain is $\Omega\backslash \calM$ and 
\begin{align}
F(\x)=\lim_{h\rightarrow 0} \frac{D}{h}\Theta(h-\mbox{dist}(\x,\partial \calM))= D\int_{\partial \calM}\delta(\x-\x')d\x'.
\end{align}
Equation (\ref{calP}a) becomes
\begin{align}
 \frac{\partial P(\x,\ell,t|\x_0)}{\partial t}&=D\nabla^2 P(\x,\ell,t|\x_0)\\
&\quad -D\int_{\partial \calM}\left (\frac{\partial P}{\partial \ell}(\x',\ell,t|\x_0) +\delta(\ell)P(\x,0,t|\x_0) \right )\delta(\x-\x')d\x',\nonumber
\end{align}
which leads to the local time BVP
\begin{subequations}
\label{Ploc}
\begin{align}
 &\frac{\partial P(\x,\ell,t|\x_0)}{\partial t}=D\nabla^2 P(\x,\ell,t|\x_0),\ \x \in \Omega\backslash \calM, \ \nabla P(\x,\ell,t|\x_0) \cdot \n_1 =0, \  \x\in \partial \Omega,\\
 &-D\nabla P(\x,\ell,t|\x_0) \cdot \n_2= D P(\x,\ell=0,t|\x_0) \ \delta(\ell)  +D\frac{\partial}{\partial \ell} P(\x,\ell,t|\x_0),  \x\in \partial \calM.
\end{align}
The unit normal $\n_2$ on $\partial \calM$ is directed towards the interior of $\calM$, see Fig. \ref{fig1}(b).
These equations are supplemented by the ''initial conditions'' $P(\x,\ell,0|\x_0)=\delta(\x-\x_0)\delta(\ell)$ and
\begin{equation}
P(\x,\ell=0,t|\x_0)=-\nabla p_{\infty}(\x,t|\x_0)\cdot \n_2 \mbox{ for } \x\in \partial \calM, 
\end{equation}
\end{subequations}
where $p_{\infty}$ is the probability density in the case of a totally absorbing surface $\partial \calM$:
\begin{subequations} 
\label{pinf}
\begin{align}
 	&\frac{\partial p_{\infty}(\x,t|\x_0)}{\partial t} = D\nabla^2 p_{\infty}(\x,t|\x_0), \, \x\in \Omega\backslash \calM,\  \nabla p_{\infty}(\x,t|\x_0) \cdot \n_1=0, \ \x\in\partial \Omega,\\
 &p_{\infty}(\x,t|\x_0)=0,\  \x\in \partial \calM,\ p_{\infty}(\x,0|\x_0)=\delta(\x-\x_0).
	\end{align}
	\end{subequations} 
One way to establish (\ref{Ploc}c) is to note that Laplace transforming equations (\ref{Ploc}a,b) with respect to $\ell$ leads to a Robin BVP, see section 3.

In the case of the occupation time (\ref{occ}), the bounded domain is $\Omega$ and
\begin{equation}
F(\x)=I_{\calM}(\x)=\int_{\calM}\delta(\x-\x')d\x'.
\end{equation}
 Equations (\ref{calP}) becomes
\begin{align}
 \frac{\partial P(\x,a,t|\x_0)}{\partial t}&=D\nabla^2 P(\x,a,t|\x_0)\\
 &\quad -\int_{\calM}\left (\frac{\partial P}{\partial a}(\x',a,t|\x_0) +\delta(a)P(\x',0,t|\x_0) \right )\delta(\x-\x')d\x'\nonumber
\label{Pocc}
\end{align}
for all $\x \in \Omega$, together with the Neumann boundary condition on $\partial \Omega$. That is,
\begin{subequations}
\label{Pocc}
\begin{align}
 \frac{\partial P(\x,a,t|\x_0)}{\partial t}&=D\nabla^2 P(\x,a,t|\x_0), \ \x \in \Omega\backslash \calM,\\
 \nabla P(\x,a,t|\x_0) \cdot \n_1&=0, \  \x \in \partial \Omega, \\
 \frac{\partial Q(\x,a,t|\x_0)}{\partial t}&=D\nabla^2 Q(\x,a,t|\x_0) -\left (\frac{\partial Q}{\partial a}(\x,a,t|\x_0) +\delta(a)Q(\x,0,t|\x_0) \right )
\end{align}
for $ \x \in \calM$,
where the propagator within $\calM$ is denoted by $Q$.
We also have the continuity conditions
	\begin{equation}
	P(\x,a,t|\x_0)=Q(\x,a,t|\x_0),\quad \nabla Q(\x,a,t|\x_0)\cdot \n_2 =\nabla P(\x,a,t|\x_0) \cdot \n_2 ,\   \x \in \partial \calM,
	\end{equation}
	\end{subequations}
	and the initial conditions $P(\x,a,0|\x_0)=\delta(\x-\x_0)\delta(a)$, $Q(\x,a,0|\x_0)=0$. We assume that the particle starts out in the non-absorbing region. (The analysis is easily modified if $\x_0\in \calM$.)
		
Given the solution to the appropriate propagator BVP, we can introduce a probabilistic model of partial absorption by generalizing the encounter-based formulation of diffusion-mediated surface reactions \cite{Grebenkov20}. That is, given the local or occupation time $\calU_t$, introduce the stopping time condition
\begin{equation}
\label{TA}
{\mathcal T}=\inf\{t>0:\ \calU_t >\widehat{\calU}\},
\end{equation}
where $\widehat{\calU}$ is a random variable with probability distribution $\Psi(u)$. Heuristically speaking, ${\mathcal T}$ is a random variable that specifies the time of absorption on $\partial \calM$ (or within $\calM$), which is the event that $\calU_t$ first crosses a randomly generated threshold $\widehat{\calU}$. The marginal probability density for particle position $\X_t $ is then
\[p(\x,t|\x_0)d\x=\P[\X_t \in (\x,\x+d\x), \ t < {\mathcal T}|\X_0=\x_0].\]
Given that $\calU_t$ is a nondecreasing process, the condition $t < {\mathcal T}$ is equivalent to the condition $\calU_t <\widehat{\calU}$. This implies that 
\begin{align*}
p(\x,t|\x_0)d\x&=\P[\X_t \in (\x,\x+d\x), \ \calU_t < \widehat{\calU}|\X_0=\x_0]\\
&=\int_0^{\infty} du\,  \psi(u)\P[\X_t \in (\x,\x+d\x), \ \calU_t < u |\X_0=\x_0]\\
&=\int_0^{\infty} du \ \psi(u)\int_0^{u} du' [P(\x,u',t|\x_0)d\x].
\end{align*}
where $\psi(u)=-d\Psi(u)/du$. Using the identity
\[\int_0^{\infty}du\ f(u)\int_0^u du' \ g(u')=\int_0^{\infty}du' \ g(u')\int_{u'}^{\infty} du \ f(u)\]
for arbitrary integrable functions $f,g$, it follows that
\begin{align}
\label{peep}
p(\x,t|\x_0)&=\int_0^{\infty}\Psi(u) P(\x,u,t|\x_0)du.
\end{align}

One general quantity of interest is the survival probability $S(\x_0,t)$ that the particle hasn't been absorbed up to time $t$, given that it started at $\x_0$. In the case of a partially absorbing surface $\partial \calM$, we have
\begin{equation}
S(\x_0,t)=\int_{\Omega\backslash \calM}p(\x,t|\x_0)d\x.
\end{equation}
Differentiating both sides with respect to time and using equations (\ref{Ploc}) shows that
\begin{align}
\frac{\partial S(\x_0,t)}{\partial t}&=D\int_{\Omega\backslash \calM} \nabla^2 p(\x,t|\x_0)d\x =-\int_{\partial \calM} \int_0^{\infty} \psi(a) P(\x,\ell,t|\x_0)da\ d\x \nonumber \\
&\equiv-J(\x_0,t),
\end{align}
where $J(\x_0,t)$ is the total probability flux into $\partial \calM$. Similarly, in the case of a partially absorbing substrate $\calM$,
\begin{equation}
\label{Socc}
S(\x_0,t)= \int_{\Omega\backslash \calM} p(\x,t|\x_0)d\x+\int_{ \calM}  q(\x,t|\x_0)d\x,
\end{equation}
and
\begin{align}
\frac{\partial S(\x_0,t)}{\partial t}&=D\int_{\Omega\backslash \calM} \nabla^2 p(\x,t|\x_0)d\x+D \int_{  \calM}\nabla^2 q(\x,t|\x_0)d\x\nonumber \\
&\quad -\int_{\calM} \int_0^{\infty} \psi(a) Q(\x,a,t|\x_0)da\ d\x.
\label{Qocc}
\end{align}
Applying the divergence theorem to the first two integrals on the right-hand side, imposing the Neumann boundary condition on $\partial \Omega$ and flux continuity at $\partial \calM$ shows that these two integrals cancel. The result is then
\begin{eqnarray}
\frac{\partial S(\x_0,t)}{\partial t}=-\int_{\calM} \int_0^{\infty} \psi(a) Q(\x,a,t|\x_0)da\ d\x =-J(\x_0,t),
\end{eqnarray}
where $J(\x_0,t)$ is now the probability flux due to absorption within the target domain $\calM$. 
Finally, Laplace transforming equation (\ref{Socc}) or (\ref{Qocc}) with respect to $t$ and noting that $S(\x_0,0)=1$ gives
\begin{equation}
\label{QL}
s\widetilde{S}(\x_0,s)-1=- \widetilde{J}(\x_0,s)
\end{equation}
with $\widetilde{f}(t)\equiv \int_0^{\infty } f(t)\e^{-st}dt$.
The probability density of the stopping time $\calT$, equation (\ref{TA}), is given by $-\partial S/\partial t$ so that the MFPT (if it exists) is
\begin{align}
\label{MFPT1}
T(\x_0)&=-\int_0^{\infty}t\frac{\partial S(\x_0,t)}{\partial t}dt =\int_0^{\infty} S(\x_0,t)dt =\S(\x_0,0)=-\left .\frac{\partial \widetilde{J}(\x_0,s)}{\partial s}\right |_{s=0}.
\end{align}
Similarly, higher order moments of the FPT density can be obtained in terms of higher order derivatives of $\widetilde{J}(\x_0,s)$. This analysis suggests that it is convenient to solve the propagator BVPs in Laplace space. It turns out that considerable simplification occurs if we consider double Laplace transforms along the lines of equation (\ref{dLT}).

\setcounter{equation}{0}

  \section{Laplace transformed BVP}
	
A crucial element of the encounter-based approach is that when $\Psi(u)=\e^{-zu}$ in equation (\ref{peep}), the resulting marginal probability density is equivalent to the Laplace transform of the generalized propagator with respect to $z$. In particular, the corresponding Laplace transformed BVP reduces to a classical form. 

\subsection{Local time propagator}

Laplace transforming the local time BVP (\ref{Ploc}) with respect to $\ell$ and setting
\begin{equation}
\widetilde{P}(\x,z,t|\x_0)=\int_0^{\infty}\e^{-z \ell}P(\x,\ell,t|\x_0)d\ell
\end{equation}
yields
\begin{subequations}
\label{PlocLT0}
\begin{align}
 &\frac{\partial \PP(\x,z,t|\x_0)}{\partial t}=D\nabla^2 \PP(\x,z,t|\x_0),\ \x \in \Omega\backslash \calM, \ \nabla \PP(\x,z,t|\x_0) \cdot \n_1 =0,\ \x\in \partial \Omega,\\
 &-\nabla \PP(\x,z,t|\x_0) \cdot \n_2= z \PP(\x,z,t|\x_0),\  \x\in \partial \calM,
\end{align}
\end{subequations}
and $\PP(\x,z,0|\x_0)=\delta(\x-\x_0)$.
We see that equation (\ref{PlocLT0}b) is a classical Robin boundary condition on $\partial \calM$ with an effective constant reactivity $\kappa_0 =zD$. Hence, as previously shown in Ref. \cite{Grebenkov20}, the Robin boundary condition is equivalent to an exponential law for the stopping local time $\widehat{\ell}_t$. Moreover,
suppose that the Robin boundary condition is rewritten as
\begin{align}
 \nabla \widetilde{P}(\x,z,t|\x_0)\cdot \n_2&=-z \widetilde{P}(\x,z,t|\x_0)=-z \int_0^{\infty}\e^{-z \ell}P(\x,\ell,t|\x_0)d\ell , \ \x \in \partial \calM.
\end{align}
Taking the limit $z \rightarrow \infty$ on both sides with $\widetilde{P}(\x,z,t|\x_0)\rightarrow p_{\infty}(\x,t|\x_0)$, and noting that $\lim_{z \rightarrow \infty}z \e^{-z\ell}$ is the Dirac delta function on the positive half-line, we obtain the supplementary condition (\ref{Ploc}c). Given the solution to the Robin BVP, we can then introduce a more general probability distribution $\Psi(\ell) = \P[\ellh>\ell]$ for the stopping local time $\ellh$ such that \cite{Grebenkov19b,Grebenkov20,Grebenkov22}
  \begin{equation}
  \label{oo}
  p (\x,t|\x_0)=\int_0^{\infty} \Psi(\ell){\mathcal L}_{\ell}^{-1}[\PP(\x,z,t|\x_0)]d\ell ,\   \x \in \Omega\backslash \calM .
  \end{equation}
This accommodates a much wider class of surface reactions where, for example, the reactivity $\kappa(\ell)$ depends on the local time $\ell$ (or the number of surface encounters):
\begin{equation}
\label{kaell}
\Psi(\ell)=\exp\left (-\frac{1}{D}\int_0^{\ell}\kappa(\ell')d\ell'\right ).
\end{equation}
It follows that one should express the solution of the Robin BVP in form that is convenient for evaluating the inverse Laplace transform, see section 4 and \cite{Grebenkov20}.

As we highlighted at the end of section 2, it is convenient to consider the BVP for the double Laplace transform (\ref{dLT}), which in the case of the local time takes the form \cite{Grebenkov20,Grebenkov22}
\begin{subequations}
\label{PlocLT}
\begin{align}
 &D\nabla^2 \calP(\x,z,s|\x_0)-s\calP(\x,z,s|\x_0)=-\delta(\x-\x_0),\ \x \in \Omega\backslash \calM,\\ & \nabla \calP(\x,z,s|\x_0)\cdot \n_1=0,\ \x \in \partial \Omega,\\
&-\nabla \calP(\x,z,s|\x_0) \cdot \n_2=z\calP(\x,z,s|\x_0) ,\ \x\in \partial \calM .
\end{align}
\end{subequations}
Laplace transforming equation (\ref{oo}) with respect to $t$ gives
  \begin{equation}
  \label{oo2}
  \p  (\x,s|\x_0)=\int_0^{\infty} \Psi(\ell){\mathcal L}_{\ell}^{-1}[\calP(\x,z,s|\x_0)] d\ell, \   \x \in \Omega\backslash \calM .
  \end{equation}
There are various quantities of interest that can be obtained directly from $ \p(\x,s|\x_0)$ without having to convert back to the time domain. For example, the MFPT for absorption is given by equation (\ref{MFPT1}) with
\begin{equation}
\label{JLT}
\widetilde{J}(\x_0,s)=D\int_{\partial \calM}\left [\int_0^{\infty}\psi(\ell){\mathcal L}_{\ell}^{-1}[\calP(\x,z,s|\x_0)] d\ell \right ]d\x.
\end{equation}

\subsection{Occupation time propagator}

Laplace transforming the occupation time BVP (\ref{Pocc}) with respect to $a$ and setting
\begin{align}
\PP(\x,z,t|\x_0)&=\int_0^{\infty} \e^{-za} P(\x,a,t|\x_0)da,\, \Q(\x,z,t|\x_0)=\int_0^{\infty} \e^{-za} Q(\x,a,t|\x_0) da,
\end{align}
yields
\begin{subequations}
\label{PoccLT}
\begin{align}
 \frac{\partial \PP(\x,z,t|\x_0)}{\partial t}&=D\nabla^2 \PP(\x,z,t|\x_0) , \ \x \in \Omega\backslash \calM,\\
 \nabla  \PP(\x,z,t|\x_0)\cdot \n_1&=0,\ \x \in \partial \Omega, \\
 \frac{\partial \Q(\x,z,t|\x_0)}{\partial t}&=D\nabla^2 \Q(\x,z,t|\x_0) -z \Q(\x,z,t|\x_0),\  \x \in \calM,\\
 \PP(\x,z,t|\x_0)&=\Q(\x,z,t|\x_0),\quad \nabla \PP(\x,z,t|\x_0)\cdot \n_2 =\nabla \Q(\x,z,t|\x_0) \cdot \n_2 
\end{align}
for $\x \in \partial \calM$.
	\end{subequations}
This is a classical BVP for diffusion in a domain with a partially absorbing substrate $\calM$ with a constant rate of absorption $z$. Note that $z$ has units of inverse time for the occupation time propagator, whereas it has units of inverse length in the case of the local time propagator. 
Following along similar lines to the local time BVP, we 
take a double Laplace transform with respect to both $t$ and $a$ by setting
\begin{subequations}
\begin{align}
\calP(\x,z,s|\x_0)&=\int_0^{\infty} \e^{-za}\left [\int_0^{\infty} \e^{-st}P(\x,a,t|\x_0)dt\right ]da,\\ \calQ(\x,z,s|\x_0)&=\int_0^{\infty} \e^{-za}\left [\int_0^{\infty} \e^{-st}Q(\x,a,t|\x_0)dt\right ]da,
\end{align}
\end{subequations}
This yields
\begin{subequations}
\label{PoccdLT}
\begin{align}
& D\nabla^2 \calP(\x,z,s|\x_0)-s\calP(\x,z,s|\x_0)=-\delta(\x-\x_0),\ \x \in \Omega\backslash \calM , \\
&-\nabla \calP(\x,z,s|\x_0) \cdot \n_1=0,\  \x\in \partial \Omega,\\
 &D\nabla^2 \calQ(\x,z,s|\x_0)-(s+z)\calQ(\x,z,s|\x_0) =0,\, \x\in \calM, \\
	&\calP(\x,z,s|\x_0)=\calQ(\x,z,s|\x_0),\ \nabla \calP(\x,z,s|\x_0)\cdot \n_2 =\nabla \calQ(\x,z,s|\x_0) \cdot \n_2, \   \x \in \partial \calM.
	\end{align}
	\end{subequations} 		
Again, given the solution to equations (\ref{PoccdLT}), we can introduce a more general probability distribution $\Psi(a) $ for the stopping occupation time such that 
 \begin{subequations}
   \label{coo}
  \begin{align}
  \p (\x,s|\x_0)&=\int_0^{\infty} \Psi(a){\mathcal L}_a^{-1}[\calP(\x,z,s|\x_0)]da,\  \x \in \Omega\backslash \calM, \\
  \q (\x,s|\x_0)&=\int_0^{\infty} \Psi(a){\mathcal L}_a^{-1}[\calQ(\x,z,s|\x_0)]da,\  \x \in \calM .
  \end{align}
  \end{subequations}
  Moreover, the MFPT for absorption within $\calM$ is determined from equation (\ref{MFPT1}) with  
\begin{equation}
\label{JLT2}
\widetilde{J}(\x_0,s)=D\int_{\calM}\left [\int_0^{\infty}\psi(a){\mathcal L}_{a}^{-1}[\calQ(\x,z,s|\x_0)] da \right ]d\x.
\end{equation}

\begin{figure}[t!]
  \centering
  \begin{tikzcd}[row sep=huge]
&\mbox{BVP with constant reactivity z} \arrow[d,"{\mathcal L}_s"] \\ P(\x,u,t|\x_0)  \arrow[r,"{\mathcal L}_z\cdot {\mathcal L}_s"]&\calP(\x,z,s|\x_0) \arrow[d,"{\mathcal L}_{u}^{-1}"] \\ &\PP(\x,u,s|\x_0) \arrow [r,"\Psi"]  & \p(\x,s|\x_0)
  \end{tikzcd}
  \caption{Commutative diagram illustrating how to incorporate the solution to a classical BVP with constant reactivity $z$ into a more general theory of diffusion in partially absorbing media. This involves a propagator $P(\x,u,t|\x_0)$ and a stopping local time distribution $\Psi(u)$.}
  \label{fig2}
  \end{figure}
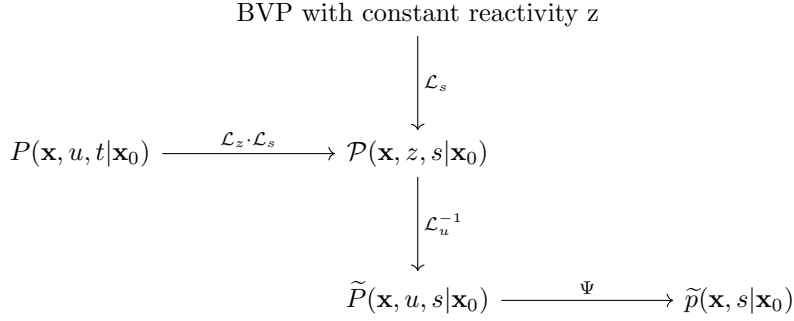

The general probabilistic framework for analyzing single-particle diffusion in partially absorbing media is summarized in the commutative diagram of Fig. \ref{fig2}. One of the challenges of implementing this method is that solutions of the classical BVPs with a constant reactivity $z$ tend to have a non-trivial parametric dependence on the Laplace variable $z$, which makes it difficult to calculate the inverse transform. In the case of reactive surfaces, solving the Robin BVP in terms of the spectrum of an associated Dirichlet-to-Neumann operator yields a series expansion that is easily inverted with respect to the Laplace variable $z$ conjugate to the local time $\ell$ \cite{Grebenkov20,Grebenkov22}. In the next section we apply this approach to the occupation time propagator.

\setcounter{equation}{0}

\section{Spectral decomposition of the occupation time propagator}

It is well known from classical PDE theory that the solution of a general Robin BVP can be computed in terms of the spectrum of a Dirichlet-to-Neumann operator. This was applied to the single-particle local time propagator BVP in Ref. \cite{Grebenkov20}. Here we carry out an analogous procedure for the occupation time BVP (\ref{PoccLT}). The basic idea is to replace the matching conditions (\ref{PoccLT}d) by the inhomogeneous Dirichlet condition $\calP(\x,s|\x_0)=\calQ(\x,s|\x_0)=f(\x,s)$ for all $\x\in \partial \calM$ and to find the function $f$ for which $\calP$ and $\calQ$ are also the solution to the original BVP. (For the moment we drop the explicit dependence of the solutions on $z$.) Therefore, consider the modified BVP
\begin{subequations}
\label{mzLT}
\begin{align}
& D\nabla^2 \calP(\x,s|\x_0)-s\calP(\x,s|\x_0)=-\delta(\x-\x_0),\ \x \in \Omega\backslash \calM , \\
&-\nabla \calP(\x,s|\x_0) \cdot \n_1=0,\  \x\in \partial \Omega,\\
 &D\nabla^2 \calQ(\x,s|\x_0)-(s+z)\calQ(\x,s|\x_0) =0,\, \x\in \calM \\
& \calP(\x,s|\x_0)=\calQ(\x,s|\x_0)=f(\x,s), \ \x \in \partial \calM.
 \end{align}
 \end{subequations}
with the function $f(\x,s)$ to be determined. This decouples the BVPs in the exterior and interior of $\calM$. Given the solution to equations (\ref{mzLT}) we then determine the function $f$ for which the solution also satisfies the original BVP (\ref{PoccLT}). 

The general solution of equations (\ref{mzLT}a,c) is of the form
 \begin{subequations}
  \label{sool}
 \begin{align}
 \calP(\x,s|\x_0)&= \calF_1(\x,s)+G_1(\x,s|\x_0),\ \x \in \Omega \backslash \calM,\\  \calQ(\x,s|\x_0)&= \calF_2(\x,s),\ \x \in \calM,
 \end{align}
 \end{subequations}
 where 
 \begin{align}
  \calF_1(\x,s)&=-D\int_{\partial \calM} \partial_{\sigma'} G_1(\x',s|\x)f(\x',s)d\x',\\ \calF_2(\x,s)&=D\int_{\partial \calM} \partial_{\sigma'} G_2(\x',s+z|\x)f(\x',s)d\x',
  \end{align}
  and $G_{1,2}$ are modified Helmholtz Green's functions: 
   \begin{subequations}
   \label{nabG}
  \begin{align}
 &D\nabla^2 G_1(\x,s|\x')-sG_1(\x,s|\x')=-\delta(\x-\x'),\ \x,\x' \in \Omega\backslash \calM,\\
 &G_1(\x,s|\x')=0,\ \x\in \partial \calM, \quad \nabla G_1(\x,s|\x')\cdot \n_1=0,\ \x\in \partial \Omega,
 \end{align}
 and
\begin{align}
& D\nabla^2 G_2(\x,s|\x')-sG_2(\x,s|\x')=-\delta(\x-\x'),\ \x,\x' \in \calM , \\
&G_2(\x,s|\x')=0,\ \x \in \partial \calM.
\end{align}
\end{subequations}
The Green's functions have dimensions of [time]/[Length]$^{d}$
The unknown function $f$ is determined by substituting the solutions (\ref{sool}b) into equation (\ref{PoccLT}b):
\begin{align}
\label{fL}
\L_s[f](\x,s)+\partial_{\sigma}G_1(\x,s|\x_0)=-\overline{\L}_{s+z}[f](\x,s),\quad \x \in \partial \calM,
\end{align}
where $\L_s$ and $\overline{\L}_s$ are the Dirichlet-to-Neumann operators
\begin{subequations}
\label{DtoN}
\begin{align}
\L_s[f](\x,s) &=-D\partial_{\sigma}\int_{\partial \calM}\partial_{\sigma'}G_1(\x',s|\x)f(\x',s)d\x',\\ \overline{\L}_s[f](\x,s) &=-D\partial_{\sigma}\int_{\partial \calM}\partial_{\sigma'}G_2(\x',s|\x)f(\x',s)d\x'.
\end{align}
\end{subequations}
acting on the space $L_2(\partial \calM)$. In the above equations we have set  $\partial_{\sigma}=\n_2\cdot \nabla_{\x}$ and  $\partial_{\sigma'}=\n_2\cdot \nabla_{\x'}$.

When the surface $\partial \calM$ is bounded, the Dirichlet-to-Neumann operators $\L_s$ and $\overline{\L}_s$ have discrete spectra. That is, there exist countable sets of eigenvalues $\mu_n(s),\overline{\mu}_n(s)$ and eigenfunctions $v_n(\x,s),\overline{v}_n(s)$ satisfying (for fixed $s$)
\begin{equation}
\label{eig}
\L_s v_n(\x,s)=\mu_n(s)v_n(\x,s),\quad \overline{\L}_s \overline{v}_n(\x,s)=\overline{\mu}_n(s)\overline{v}_n(\x,s).
\end{equation}
It can be shown that the eigenvalues are non-negative and that the eigenfunctions form a complete orthonormal basis in $L_2(\partial \Omega_2)$. We can now solve equation (\ref{fL}) by introducing an eigenfunction expansion of $f$ with respect to one of the operators. For concreteness, we take
\begin{equation}
\label{eig2}
f(\x,s)=\sum_{m=0}^{\infty}f_m(s) v_m(\x,s).
\end{equation}
Substituting equation (\ref{eig2})
into (\ref{fL}) and taking the inner product with the adjoint eigenfunction $v_n^*(\x,s)$ yields the following matrix equations for the coefficients $f_m$:
\begin{equation}
\label{sspec0}
\mu_n(s)f_n(s) =g_n(s)-\sum_{m\geq 1} H_{nm}(s+z)f_m(s),
\end{equation}
where
\begin{align}
g_n(s)&=-\int_{\partial \calM}v_n^*(\x,s) \partial_{\sigma}G_1(\x,s|\x_0)d\x,\\
H_{nm}(s)&=-D\int_{\partial \calM} v_n^*(\x,s)\partial_{\sigma}\left \{\int_{\partial \calM}v_m(\x',s)\partial_{\sigma'}G_2(\x',s|\x) d\x'\right \}d\x.
\label{H}
\end{align}
The orthogonality condition 
\begin{equation}
\int_{\partial \calM} v_n^*(\x,s)v_m(\x,s)d\x=\delta_{m,n}
\end{equation}
means that $v_n^*$ and $v_m$ can each be taken to have dimensions of [Length]$^{-(d-1)/2}$. It also follows that $H_{nm}(s)$ has dimensions of inverse length.

Introducing the vectors ${\bf f}(s)=(f_n(s), n\geq 0)$ and ${\bf g}(s)=(g_n(s), n\geq 0)$, we can formally write the solution of equation (\ref{sspec0}) as
\begin{equation}
\label{eff}
{\bf f}(s)=\left [{\bf M}(s)+{\bf H}(s+z)\right ]^{-1} {\bf g}(s),
\end{equation}
where ${\bf H}(s)$ is the matrix with elements $H_{nm}(s)$ and ${\bf M}(s)=\mbox{diag}(\mu_1(s),\mu_2(s)\ldots)$. Finally, substituting equation (\ref{eff}) into equations (\ref{sool}) gives
\begin{subequations}
 \label{spec1}
 \begin{align}
 \calP(\x,z,s|\x_0)&=G_1(\x,s|\x_0)\\
 &\quad +\frac{1}{D}\sum_{n,m}{\mathcal V}_n(\x,s)\left [{\bf M}(s)+{\bf H}(s+z)\right ]_{nm}^{-1}{\mathcal V}^*_m(\x_0,s), \x \in \Omega \backslash \calM,\nonumber \\  \calQ(\x,z,s|\x_0)&=\frac{1}{D} \sum_{n,m}\widehat{\mathcal V}_n(\x,s+z)\left [{\bf M}(s)+{\bf H}(s+z)\right ]_{nm}^{-1}{\mathcal V}^*_m(\x_0,s),\ \x \in \calM,
 \end{align}
 \end{subequations}
 where 
 \begin{subequations}
 \label{Vn}
 \begin{align}
{\mathcal V}_n(\x,s)&=-D\int_{\partial \calM} v_n(\x',s)\partial_{\sigma'}G_1(\x',s|\x)d\x',\\
\widehat{\mathcal V}_n(\x,s)&=D\int_{\partial \calM} v_n(\x',s)\partial_{\sigma'}G_2(\x',s|\x)d\x'.
\end{align}  
\end{subequations}

An analogous construction can be carried out for the local time BVP (\ref{PlocLT}) by decomposing the generalized propagator as \cite{Grebenkov20}
\begin{equation}
\calP(\x,z,s|\x_0)=G_1(\x,s|\x_0)+\calF(\x,z,s|\x_0),
\end{equation}
with
\begin{subequations}
\label{homT}
\begin{align}
 &D\nabla^2 \calF(\x,z,s|\x_0)-s\calF(\x,z,s|\x_0)=0,\ \x \in \Omega\backslash \calM,\\
&\nabla \calF(\x,z,s|\x_0) \cdot \n_2+z\calF(\x,z,s|\x_0) =- \nabla G_1(\x,s|\x_0) \cdot \n_2\mbox{ for }  \x\in \partial \calM, \\
&D\nabla \calF(\x,z,s|\x_0) \cdot \n_1=0 \mbox{ for }  \x\in \partial \Omega.
\end{align}
\end{subequations} 
Replacing the Robin condition by the Dirichlet condition $\calF(\x,z,s|\x_0)=f(\x,s)$ leads to the equation
\begin{equation}
\L_s[f](\x,s)+zf(\x,s)=-\partial_{\sigma}G_1(\x,s|\x_0),
\end{equation}
where $\L_s$ is the Dirichlet-to-Neumann operator (\ref{DtoN}a).
Again this can be solved by substituting for $f$ using the eigenfunction expansion (\ref{eig2}), which yields the result
\begin{equation}
 \label{spec2}
\calP(\x,z,s|\x_0)=G_1(\x,s|\x_0)+\frac{1}{D}\sum_{n=0}^{\infty} \frac{{\mathcal V}_n(\x,s){\mathcal V}^*_n(\x_0,s)}{\mu_n(s)+z},
\end{equation}
Comparison of equations (\ref{spec1}) and (\ref{spec2}), establishes that the occupation time propagator is a much more complicated function of the Laplace variable $z$. This means that, in general, the inverse Laplace transform has to be determined by computing a corresponding Bromwich integral. This, in turn, requires finding the roots of the characteristic equation $\mbox{det}[{\bf M}(s) +{\bf H}(s+z)]=0$ in order to identify the poles in the complex $z$-plane. (An analogous issue arises in solving a classical Robin BVP with a space-dependent reactivity $\kappa(\x)$, $\x \in \partial \calM$ \cite{Grebenkov19b}.) On the other hand, it is straightforward to obtain the inverse Laplace transform of equation (\ref{spec2}), assuming that we can invert term-by-term in the infinite sum. In particular \cite{Grebenkov20},
 \begin{equation}
 \label{speckle}
  \PP(\x,\ell,s|\x_0)=G_1(\x,s|\x_0)\delta(\ell)+\frac{1}{D}\sum_{n=0}^{\infty}  {\mathcal V}_n^*(\x_0,s){\mathcal V}_n(\x,s) \e^{-\mu_n(s) \ell}.
  \end{equation}

\subsubsection*{Partially absorbing sphere}

One example where the spectral decompositions of the Dirichlet-to-Neumann operators $\L_s$ and $\overline{\L}_s$ are known exactly is a partially absorbing sphere. Let $\Omega = \R^3$ and $\calM=\{\x\in \R^3,\, 0 <  |\x| <R\}$ so that $\partial \calM= \{\x\in \R^3,\,  |\x| =R\}$. (In the context of partially absorbing surfaces, $\L_s$ is the relevant operator for diffusion exterior to the sphere, whereas $\overline{\L}_s$ is the appropriate operator for diffusion within the sphere \cite{Grebenkov19b}.) The rotational symmetry of $\calM$ means that if $\L_s$ and $\overline{\L}_s$ are expressed in spherical polar coordinates $(r,\theta,\phi)$, then the eigenfunctions are given by spherical harmonics, and are independent of the Laplace variable $s$ and the radius $r$:
\begin{equation}
v_{nm}(\theta,\phi)=\overline{v}_{nm}(\theta,\phi)=\frac{1}{R} Y_n^m(\theta,\phi),\quad n\geq 0, \ |m|\leq n.
\end{equation}
From orthogonality, it follows that the adjoint eigenfunctions are
\begin{equation}
v^*_{nm}(\theta,\phi)=\overline{v}_{nm}^*(\theta,\phi)=(-1)^m\frac{1}{R} Y_n^{-m}(\theta,\phi).
\end{equation}
(Note that eigenfunctions are labeled by the pair of indices $(nm)$.)
The corresponding eigenvalues are \cite{Grebenkov19b}
\begin{equation}
\mu_n(s)=-\alpha\frac{k_n'(\alpha(s) R)}{k_n(\alpha(s) R)},\quad \overline{\mu}_n(s)=\alpha(s)\frac{i_n'(\alpha(s) R)}{i_n(\alpha(s) R)},
\end{equation}
where $\alpha(s)=\sqrt{s/D}$.
Since the $n$th eigenvalue is independent of $m$, it has a multiplicity $2n+1$.
It is also possible to compute the projections of the boundary fluxes in (\ref{Vn}) by using appropriate series expansions of the corresponding Green's functions. For example, one finds that \cite{Grebenkov20}
\begin{equation}
-D\partial_{\sigma}G_1(\x',s|\x)=\sum_{n=0}^{\infty} \frac{2n+1}{4\pi R^2} P_n(\x'\cdot \x/(rR))\frac{k_n(\alpha(s) r)}{k_n(\alpha(s) R)}
\end{equation}
with $|\x'|=R$, $|\x|=r > R$, and $P_n(x)$ a Legendre polynomial. Hence, using $\partial_{\sigma'}=-\partial/\partial r'$, we have
 \begin{align}
{\mathcal V}_{nm}(\x,s)&\equiv D\int_{|\x'|=R_1} v_{nm}(\theta',\phi')\frac{\partial}{\partial r'}G_1(\x',s|r,\theta,\phi)d\x'\nonumber \\
&=-v_{nm}(\theta,\phi)\frac{k_n(\alpha(s) r)}{k_n(\alpha(s) R)}
\end{align} 
with $\x=(r,\theta,\phi)$ and $r>R$. 
Similarly, we have
 \begin{align}
\widehat{\mathcal V}_{nm}(\x,s)&\equiv -D\int_{|\x'|=R_1} v_{nm}(\theta',\phi')\frac{\partial}{\partial r'}G_2(\x',s|r,\theta,\phi)d\x'\nonumber \\
&=-v_{nm}(\theta,\phi)\frac{i_n(\alpha(s) r)}{i_n(\alpha(s) R)}
\end{align}  
for $\x=(r,\theta,\phi)$ and $r<R$.
Finally, the matrix ${\bf H}(s)$ in equation (\ref{H}) becomes
\begin{align}
H_{nm,n'm'}(s)&=-D\int_{\partial \calM} v_{nm}^*(\theta,\phi)\frac{\partial}{\partial r}\left \{\int_{\partial \calM}v_{n'm'}(\theta',\phi')\frac{\partial}{\partial r'}G_2(\x',s|\x) d\x'\right \}d\x\nonumber \\
&=\alpha(s) \frac{i'_{n'}(\alpha(s) R)}{i_{n'}(\alpha(s) R)}\left [\int_{\partial \calM} v_{nm}^*(\theta,\phi)v_{n'm'}(\theta,\phi)d\x\right ]\nonumber \\
&=\overline{\mu}_n(s)\delta_{n,n'}\delta_{m,m'} .
\label{H2}
\end{align}
That is, ${\bf H}$ is a diagonal matrix. Finally, from equation (\ref{spec1}) the generalized propagator within the sphere becomes
 \begin{align}
  &\calQ(\x,z,s|\x_0)=\frac{1}{D} \sum_{n,m} \sum_{n',m'}\widehat{\mathcal V}_{nm}(\x,s+z)\left [{\bf M}(s)+{\bf H}(s+z)\right ]_{nm,n'm'}^{-1}{\mathcal V}^*_{n'm'}(\x_0,s), \\
  &=\frac{1}{D}\sum_{n,m}v_{nm}(\theta,\phi)\frac{i_n(\alpha(s+z) r)}{i_n(\alpha(s+z) R)}\frac{1}{\mu_n(s)+\overline{\mu}_n(s+z) }\frac{k_n(\alpha(s) r_0)}{k_n(\alpha(s) R)}v_{nm}^*(\theta_0,\phi_0) , \nonumber
 \end{align}
 where $\x=(r,\theta,\phi)$ and $\x_0=(r_0,\theta_0,\phi_0)$ with $r<R$ and $r_0>R$. An analogous result holds for $ \calP(\x,s|\x_0)$.
 
 We conclude that in the case of a sphere, one can obtain explicit expressions for the doubly Laplace-transformed propagator. However, in order to incorporate a non-exponential occupation-time distribution $\Psi(a)$ for partial absorption, it is necessary to invert the Laplace transform with respect to $z$. In general, this would have to be implemented numerically. In the next section we consider a simpler example, where the inverse Laplace transform can be obtained analytically, and use this to explore how the survival probability and MFPT depend on the choice of distribution $\Psi(a)$.

\setcounter{equation}{0}
\section{Survival probability and MFPT for a partially absorbing interval}

\begin{figure}[t!]
\centering
  \includegraphics[width=10cm]{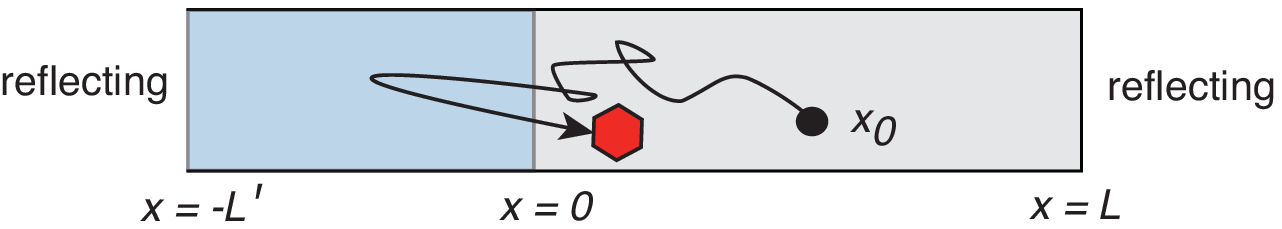}
  \caption{Partially absorbing substrate in 1D with $\calM=[-L',0]$ and $\Omega\backslash \calM=[0,L]$.}
  \label{fig3}
\end{figure}

We now consider a one-dimensional (1D) configuration for which the Dirichlet-to-Neumann operators reduce to scalars, thus greatly simplifying the analysis of the occupation time BVP. Suppose that a particle diffuses in the interval $\Omega=[-L',L]$ with a partially absorbing subinterval $\calM=[-L',0]$, see Fig. \ref{fig3}. It follows that $\partial \Omega=\{-L',L\}$ and $\partial \calM=\{0\}$. 
The 1D version of equations (\ref{PoccLT}) takes the form 
\begin{subequations}
\label{1DBVP}
\begin{align}
 & D\frac{\partial^2 \calP(x,z,s|x_0)}{\partial x^2}-s\calP(x,z,s|x_0) =- \delta(x-x_0),\ 0<x<L,  \\
  &D\frac{\partial^2 \calQ(x,z,s|x_0)}{\partial x^2}-(s+z)\calQ(x,z,s|x_0)=0,\, -L'<x<0,\\
 & \frac{\partial \calP(x,z,s|x_0)}{\partial x}=0,\ x=L,\quad  \frac{\partial \calQ(x,z,s|x_0)}{\partial x}=0,\ x=-L'.
 \end{align}
These are supplemented by the matching conditions
\begin{align}
  \calP(0,z,s|x_0)&=\calQ(0,z,s|x_0),\ \frac{\partial \calP}{\partial x}(0,z,s|x_0)  = \frac{\partial\calQ}{\partial x}(0,z,s|x_0).
\end{align}
\end{subequations}
Note that this particular BVP for fixed $z$ and $L,L'\rightarrow \infty$ was previously considered within the context of so-called virtual traps \cite{Naim93}. 
The general solution of (\ref{1DBVP}) is
\begin{subequations}
\label{sQ1D}
\begin{align}
 \calP(x,z,s|x_0)&=A(z,s)\cosh\alpha(s) (L-x) + G_1(x, s|x_0),\ x\in [0,L]\\
 \calQ(x,z,s|x_0)&=B(z,s)\cosh\alpha(s+z) (L'+x) ,\ x \in [-L',0],
\end{align}
\end{subequations}
where $\alpha(s)=\sqrt{s/D}$ and $G_1$ is the 1D Green's function that satisfies equation (\ref{1DBVP}a) with a Dirichlet boundary condition at $x = 0$
and a Neumann boundary condition at $x=L$:
\begin{align}
 G_1(x, s| x_0) 
    &= \frac{H(x_0 - x)g(x, s)\widehat{g}(x_0, s) +H(x - x_0)g(x_0, s)\widehat{g}(x, s)}{\sqrt{sD}\cosh(\sqrt{s/D}L)},
\end{align}
where $H(x)$ is the Heaviside function and
\begin{eqnarray}
   g(x, s) = \sinh \sqrt{s/D} x,\quad \makebox{and} \quad \widehat{g}(x, s) =\cosh\sqrt{s/D} (L-x). 
\end{eqnarray}

The unknown coefficients $A,B$ are determined from the matching conditions at $x= 0$, which reduce to
\begin{subequations}
\label{smatch}
\begin{align}
A(z,s)\cosh\alpha(s) L&=B(z,s)\cosh\alpha(s+z) L',\\
-\alpha(s)A(z,s)\sinh\alpha(s) L&=\alpha(s+z)B(z,s)\sinh\alpha(s+z)L' -\partial_xG_1(0,s|x_0).
\end{align}
\end{subequations}
Substituting (\ref{smatch}a) into (\ref{smatch}b) gives
\begin{align}
&\left \{\alpha(s)\tanh \alpha(s)L \cosh\alpha(s+z) L'+\alpha(s+z)\sinh\alpha(s+z) L'\right \} B(z,s) \nonumber \\
&\quad =\frac{1}{D}\frac{\cosh\sqrt{s/D} (L-x_0)}{\cosh\sqrt{s/D} L}.
\label{DN2}
\end{align}
Note that equation (\ref{DN2}) could also be derived from the 1D version of equation (\ref{sspec0}), with $\partial \calM$ corresponding to the single point $x=0$ and $\partial_{\sigma}=-\partial_x$ etc. In particular, since $G_2(x,s|x_0)$ is the Dirichlet-Neumann Green's function on $(-L',0]$, we find that 
\begin{equation}
\L_s [f](L,s)\equiv -Df(L,s)\left .\partial_{x}\partial_{x'} G_1(x',s|x)\right |_{x=x'=0}=f(0,s)\sqrt{\frac{s}{D}} \tanh(\sqrt{s/D}L) ,
\end{equation}
and
\begin{equation}
\overline{\L}_s [f](L,s)\equiv -Df(L,s)\left .\partial_{x}\partial_{x'} G_2(x',s|x)\right |_{x=x'=0}= f(0,s)\sqrt{\frac{s}{D}} \tanh(\sqrt{s/D}L').
\end{equation}
We deduce that for 1D diffusion, the Dirichlet-to-Neumann operators reduce to scalars with corresponding eigenvalues $\mu(s)=\alpha(s) \tanh (\alpha(s) L)$ and $\overline{\mu}(s)=\alpha(s) \tanh (\alpha(s) L')$.
In the specific case $x_0=0$, the solution has the particularly simple form
\begin{subequations}
\label{squid}
\begin{align}
 \calP(x,z,s|0)&=\frac{1}{\Phi(z,s)D}\frac{\cosh\alpha(s) (L-x)}{ \cosh\alpha(s) L} ,\ x\in [0,L],\\
 \calQ(x,z,s|0)&=\frac{1}{\Phi(z,s)D}\frac{\cosh\alpha(s+z) (L'+x)}{ \cosh\alpha(s+z) L'},\ x \in [-L',0],
\end{align}
\end{subequations}
where
\begin{equation}
\label{Psi}
\Phi(z,s)\equiv  \alpha(s) \tanh [\alpha(s)L] +\alpha(s+z) \tanh [\alpha(s+z)L'] .
\end{equation}

For the sake of illustration, let us focus on the behavior of the solution at the interface $x=0$ between the non-absorbing and absorbing regions, and the survival probability. Setting $x=x_0=0$ we have $\calP(0,z,s|0)=\calQ(0,z,s|0)\equiv C(z,s)$, where
\begin{equation}
C(z,s)=\frac{1}{\sqrt{D} } \frac{1}{\sqrt{s+z}\tanh(\sqrt{[s+z]/D}L')+\sqrt{s}\tanh(\sqrt{s/D}L)} .
\end{equation}
The corresponding generalized survival probability is
\begin{align}
S(z,s)&\equiv \int_{-L'}^0  \calQ(x,z,s|0)dx+\int_0^L \calP(x,z,s|0)dx\nonumber \\
&=\frac{1}{\Phi(z,s)D}\left (\frac{\tanh \alpha(s+z)L'}{\alpha(s+z)}+\frac{\tanh \alpha(s)L}{\alpha(s)}\right )\nonumber \\
&=\frac{1}{\sqrt{s(s+z)} } \frac{\sqrt{s}\tanh (\sqrt{(s+z)/D}L')+\sqrt{s+z}\tanh(\sqrt{s/D}L)}{\sqrt{s+z}\tanh  (\sqrt{(s+z)/D}L')+\sqrt{s}\tanh(\sqrt{s/D}L)}.
\label{ssz}
\end{align}
In standard treatments of partial absorption \cite{Naim93}, one simply identifies $C(\kappa_0,s)$ as the Laplace transformed probability density $\p(s)$ at the origin for a constant rate of absorption $z=\kappa_0 $ within the domain $[-L',0]$. Similarly $S(\kappa_0,s)$ is the Laplace transformed survival probability for a particle starting at $x_0=0$. It immediately follows from equations (\ref{MFPT1}) and (\ref{ssz}) that the corresponding MFPT for absorption is  
\begin{equation}
T(0)=\lim_{s\rightarrow 0} S(\kappa_0,s)= \frac{L\tanh(\sqrt{\kappa_0/D}L')}{\sqrt{\kappa_0 D}}+\frac{1}{\kappa_0}.
\end{equation}
Note that $T\rightarrow 0$  for a totally absorbing substrate ($\kappa_0\rightarrow \infty$), since the particle starts at the interface. The novel feature of the encounter-based formalism is that one can construct a more general model of absorption within $\calM= (-\infty,0]$ by inverting with respect to $z$ and introducing a stopping occupation time distribution $\Psi(a)$. The details depend on whether or not $L'$ is finite.

\subsection{Unbounded partially absorbing substrate ($L'\rightarrow \infty$)}

In the limit $L'\rightarrow \infty$ we have  $\tanh(\sqrt{(s+z)/D}L')\rightarrow 1$ and the $z$-dependence of the solutions (\ref{squid}) simplifies greatly. The effects of absorption then depend on whether or not $L$ is itself finite.

\paragraph{Limit $L\rightarrow \infty$.} First, consider the limiting case $L\rightarrow \infty$ with $\tanh(\sqrt{s/D}L)\rightarrow 1$. Using standard Laplace transform tables, we can invert with respect to $s$ or $z$:
 \begin{equation}
 \widetilde{C}(z,t)=\frac{1}{2z\sqrt{\pi D t^3}} \left (1-\e^{-zt}\right ),\quad  \widetilde{S}(z,t)=\e^{-zt/2} I_0(zt/2).
 \end{equation}
 and
  \begin{equation}
 \widetilde{C}(a,s)=
 \frac{\e^{-sa}}{\sqrt{\pi aD}}-\sqrt{\frac{s}{D}}\mbox{erfc}(\sqrt{sa}),\quad \widetilde{S}(a,s)= \frac{\e^{-sa}}{\sqrt{\pi s a}},
 \end{equation}
 where
 \begin{equation}
 \mbox{erfc}(x)=\frac{2}{\sqrt{\pi}}\int_x^{\infty}\e^{-y^2}dy.
 \end{equation}
Moreover,
\begin{equation}
C(a,t)=\frac{1}{2\sqrt{\pi D t^3}} \left (H(a)-H(a-t)\right ),\quad S(a,t)=\frac{1}{\pi \sqrt{a[t-a]}}H(t-a).
\end{equation}
For a constant rate of absorption $z=\kappa_0 $ (exponential distribution $\Psi(a)=\e^{-\kappa_0 a}$), we recover some of the results of Ref. \cite{Naim93}. For example, the density at the origin is $p(t)=\widetilde{C}(\kappa_0,t)$. Hence, at short times, $t\ll 1/\kappa_0$, the density is diffusion dominated with $p(t) \sim (4\pi D t)^{-1/2}$, whereas at large times $p(t)\sim t^{-3/2}$ such that $p(t) \rightarrow S(t)/\sqrt{D/\kappa_0}$, where $S(t)$ is the corresponding survival probability $\widetilde{S}(\kappa_0,t)$. On the other hand, for a non-exponential stopping occupation time distribution $\Psi(a)$, the marginal density at the origin becomes
\begin{equation}
p(t)=\int_0^{\infty}\Psi(a)C(a,t)da =\frac{1}{2\sqrt{\pi D t^3}} \int_t^{\infty}\Psi(a)da,
\end{equation}
and the survival probability is now
\begin{equation}
\label{surv}
S(t)=\int_0^{\infty}\Psi(a) S(a,t)da=\int_0^{t}\frac{\Psi(a)}{\pi\sqrt{a[t-a]}}da.
\end{equation}
Since both the absorbing and non-absorbing intervals are unbounded, the MFPT is infinite.

\paragraph{Finite $L$.} When $L$ is finite, one has to invert the $s$-Laplace transforms using Bromwich integrals. However, the inverse $z$-Laplace transforms are more straightforward:
\begin{equation}
 \widetilde{C}(a,s)=
 \frac{\e^{-sa}}{\sqrt{\pi aD}}-\sqrt{\frac{s}{D}}\tanh(\sqrt{s/D}L)\e^{-s\mbox{sech}^2(\sqrt{s/D}L)}\mbox{erfc}(\sqrt{sa}\tanh(\sqrt{s/D}L)).
 \end{equation}
 and
\begin{equation}
 \widetilde{S}(a,s)=
 \e^{-s\mbox{sech}^2(\sqrt{s/D}L)}\mbox{erfc}(\sqrt{sa}\tanh(\sqrt{s/D}L))+\frac{\sqrt{D}\tanh \sqrt{s/D})L}{\sqrt{s}}\widetilde{C}(a,s).
 \end{equation}
 Given a stopping occupation time distribution $\Psi(a)$, the corresponding MFPT is
 \begin{align}
 T(0)&=\lim_{s\rightarrow 0}\int_0^{\infty}\Psi(a) \widetilde{S}(a,s)da=\int_0^{\infty} \Psi(a)\left [\frac{L}{\sqrt{\pi a D}}+1 \right ]da.
 \end{align}
 Using integration by parts, we see that
\[\int_0^{\infty} \Psi(a)da=[a\Psi(a]_0^{\infty}-\int_0^{\infty}a\Psi'(a)da= \int_0^{\infty}a\psi(a)da=-\widetilde{\psi}'(0).\]
 Hence, a necessary condition for the existence of $T(0)$ is that $\psi(a)$ has a finite first moment. 
 \medskip
 
  \begin{figure}[t!]
\centering
  \includegraphics[width=11cm]{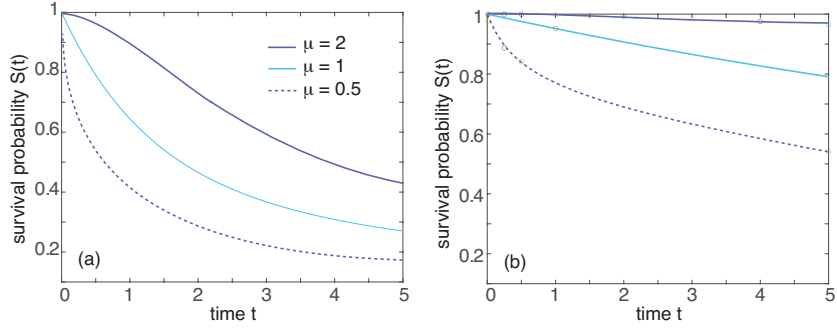}
  \caption{Plot of the survival probability $S(t) $ of equation (\ref{surv}) as a function of time $t$ in the case of the gamma distribution. (a) $\gamma=1$; (b) $\gamma=0.1$.}
  \label{fig4}
\end{figure}

 \begin{figure}[t!]
\centering
  \includegraphics[width=11cm]{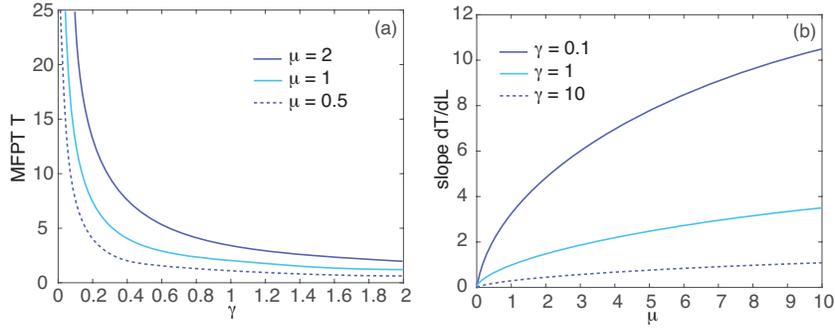}
  \caption{MFPT $T$ for the gamma distribution. (a) Plot of $T$ as a function of $\gamma$ for various values of $\mu$. The case $\mu=1$ corresponds to the exponential distribution (constant reactivity). (b) Corresponding plots of the slope $dT/dL$ as a function of $\mu$ for various values of $\gamma$.}
  \label{fig5}
\end{figure}

 One non-exponential distribution that has finite moments is the gamma distribution:
\begin{equation}
\label{psigam}
\psi_{\rm gam}(a)=\frac{\gamma(\gamma a)^{\mu-1}\e^{-\gamma a}}{\Gamma(\mu)},\quad \Psi(a)=\frac{\Gamma(\mu,\gamma a)}{\Gamma(\mu)} ,\ \mu >0,
\end{equation}
where $\Gamma(\mu)$ is the gamma function and $\Gamma(\mu,z)$ is the upper incomplete gamma function:
\begin{equation}
\Gamma(\mu)=\int_0^{\infty}\e^{-t}t^{\mu-1}dt,\quad \Gamma(\mu,z)=\int_z^{\infty}\e^{-t}t^{\mu-1}dt,\ \mu >0.
\end{equation}
The corresponding Laplace transform is
\begin{equation}
\widetilde{\psi}_{\rm gam}(z)=\left (\frac{\gamma}{\gamma+z}\right )^{\mu}.
\end{equation}
Here $\gamma$ determines the effective absorption rate so that the substrate $\calM$ is non-absorbing in the limit $\gamma\rightarrow 0$ and totally absorbing in the limit $\gamma \rightarrow \infty$. (In the latter case, if $x_0>0$ then the particle is absorbed as soon as it reaches $x=0$.) If $\mu=1$ then $\psi_{\rm gam}$ reduces to the exponential distribution with constant reactivity $\gamma$, that is, $\psi_{\rm gam}(a)|_{\mu =1}=\gamma \e^{-\gamma a}$. The parameter $\mu$ thus characterizes the deviation of $\psi_{\rm gam}(a)$ from the exponential case. If $\mu <1$ ($\mu>1$) then $\psi_{\rm gam}(a)$ decreases more rapidly (slowly) as a function of the occupation time $a$. 
In Fig \ref{fig4} we plot the survival probability $S(t)$ of equation (\ref{surv}) as a a function of time for various parameter values of the gamma distribution.
As expected, the survival probability at a given time $t$ is larger for smaller $\gamma$ or larger $\mu$. In Fig. \ref{fig5}(a) we plot the corresponding MFPT $T(0)$ as a function of $\gamma$ for various values of $\mu$. It can be seen that as $\gamma \rightarrow \infty$ all the curves converge to zero. This is a consequence of the fact that the domain $\calM=(-\infty,0]$ is totally absorbing in this limit and $x_0=0$. Fig. \ref{fig5}(b) shows that increasing $\mu$ increases the sensitivity of the MFPT to variations in the length of the non-absorbing domain. This is a particularly significant effect when the rate of absorption $\gamma$ is small.

\subsection{Bounded partially absorbing substrate ($L'<\infty$)}

\begin{figure}[b!]
\centering
  \includegraphics[width=11cm]{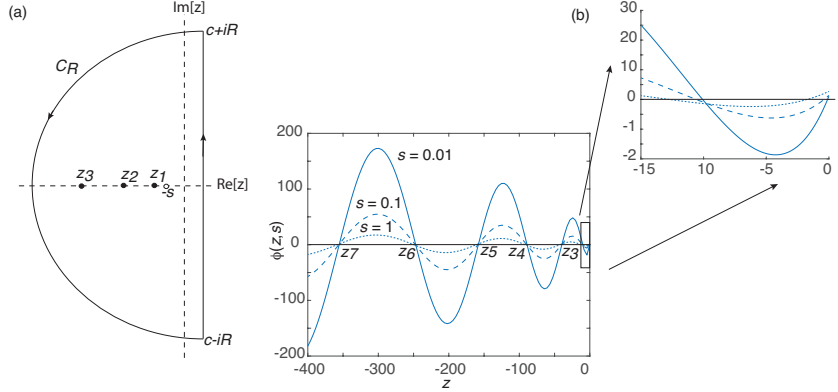}
  \caption{(a) Contour $C$ for evaluating the Bromwich integral (\ref{brom1D}). The function $\calQ(x,z,s|x_0)$ has a removable singularity at $z=-s$, and an infinite set of poles at $z=z_n <-s$ for $n\geq 1$ and $R\rightarrow \infty$. (b) Plot of the function $\phi(z,s)\equiv \Phi(z,s)/\alpha(s)$ against $z$ for various values of $s$. Inset shows curves in a neighborhood of $z=0$. We have also set $D=1$ and $L'=1$ so that $s_0=1$.}
  \label{figZ}
\end{figure}

The $z$-dependence of the solutions (\ref{squid}) is more complicated when $L'$ is finite. For the sake of illustration, let $L\rightarrow \infty$ so that the solutions (\ref{squid}) become (for $x_0=0$)
\begin{subequations}
\label{squid2}
\begin{align}
 \calP(x,z,s|0)&=\frac{\e^{-\alpha(s)x}}{\Phi(z,s)D}  ,\ x\geq 0,\\
 \calQ(x,z,s|0)&=\frac{1}{\Phi(z,s)D}\frac{\cosh\alpha(s+z) (L'+x)}{ \cosh\alpha(s+z) L'},\ x \in [-L',0],
\end{align}
\end{subequations}
where
\begin{equation}
\label{Psi2}
\Phi(z,s)\equiv  \alpha(s)   +\alpha(s+z) \tanh [\alpha(s+z)L'] .
\end{equation}
In order to find the inverse Laplace transforms we now have use Bromwich integrals. For example,
the inverse $z$-Laplace transform of $\calQ(x,z,s|0)$ takes the form
\begin{equation}
\label{brom1D}
\Q(x,a,s|x_0)=\frac{1}{2\pi i  }\int_{c-i\infty}^{c+i\infty} \e^{za}\frac{1}{D\Phi(z,s)}\frac{\cosh\alpha(s+z) (L'+x)}{ \cosh\alpha(s+z) L'} dz,
\end{equation}
with $c$, $c>0$, chosen so that the Bromwich contour is to the right of all singularities of $\calQ(x,z,s|x_0)$. 
The Bromwich integral (\ref{brom1D}) can be evaluated by closing the contour in the complex $z$-plane as illustrated in Fig. \ref{figZ}(a). The resulting contour encloses a countably infinite number of poles, which correspond to the zeros of the function $\Phi(z,s)$.\footnote{The dependence of $\calQ(x,z,s|x_0)$ on $\alpha(s+z)=\sqrt{[s+z]/D}$ suggests that there is also a branch point at $z=-s$. However, $\cosh[\alpha(s+z)L']$ and $\alpha(s+z)\sinh [\alpha(s+z)L']$ are even functions of $\alpha(s+z)$, so $z=-s$ is a removable singularity and $\calQ(x,z,s|x_0)$ is single-valued.} Setting $\Phi(z,s)=0$ in equation (\ref{Psi2}) and rearranging leads to the transcendental equation
\begin{equation}
\label{tran}
\tanh y=-\sqrt{\frac{s}{s_0}}\frac{1}{y},\quad s_0=\frac{D}{{L'}^2},\quad y=L'\sqrt{[s+z]/D}.
\end{equation}
Clearly (\ref{tran}) does not have any real solutions. However, there exists a countably infinite number of pure imaginary solutions $y=i\omega_n$, $n\geq 1$,  with $\omega_n$ real such that
\begin{equation}
\tan \omega_n = \sqrt{\frac{s}{s_0}}\frac{1}{\omega_n}.
\end{equation}
The corresponding zeroes in the $z$-plane are real and lie to the left of $z=-s$ since $\omega_n \neq 0$:
\begin{equation}
z_n=-s-s_0 \omega_n^2,\quad n\geq 1.
\end{equation}
Example plots of the function $\Phi(z,s)/\alpha(s)$ are shown in Fig. \ref{figZ}(b). Since the roots are ordered on the negative real axis, there exists an integer $N=N(s)$ such that $\omega_n\gg \sqrt{s/s_0}$ for all $n \geq N(s)$. This implies that $\tan \omega_n \approx 0$ or $\omega_n \approx n \pi $ for all $n\geq N(s)$.
In other words,
\begin{equation}
z_n\approx -s-s_0 (n\pi)^2,\quad n\geq N(s).
\end{equation}
Moreover, in the small-$s$ regime ($s\ll s_0$), we have $N(s)=O(1)$.

Applying Cauchy's residue theorem to the Bromwich contour integral of Fig. \ref{figZ}(a), and noting that the contribution from the semi-circle $C_R$ vanishes in the limit $R\rightarrow \infty$, we have
\begin{equation}
\label{brom1D2}
\Q(x,a,s|x_0)=\sum_{n\geq 1} \frac{1}{\Phi_n(s)}\e^{-(s+s_0\omega_n^2)a}\frac{\cos[\omega_n (L'+x)/L']}{\cos\omega_n} ,
\end{equation}
We have used $\alpha(z_n+s)=i\omega_n/L'$ and set 
\begin{eqnarray}
\partial_z\Phi_{\infty}(z_n,r)&=&  \frac{L'}{2D }\left (\frac{\tan \omega_n}{\omega_n} +\sec^2 \omega_n \right)\nonumber \\
&=&  \frac{L'}{2D }\left (1 +\sqrt{\frac{r}{r_0}}\left [1+\sqrt{\frac{r}{r_0}}\right ] \frac{1}{ \omega_n^2} \right),
\label{Psin2}
\end{eqnarray}Since $\omega_n\approx n\pi$ and $\Phi_n(s)\approx  L'/2D$ for $n\geq N(s)$, it follows that for small $s$ only the first few terms in the infinite sum of equation (\ref{brom1D2}) have to be computed numerically.

\section{Conclusion}  

In this paper we showed how transform methods and the spectral theory of Dirichlet-to-Neumann operators can be used to solve a general class of BVPs arising from models of single-particle diffusion in partially absorbing media. In particular, we extended the encounter-based probabilistic framework for analyzing diffusion-mediated surface absorption \cite{Grebenkov20,Grebenkov22} to the case of partially absorbing interiors. There are a number of applications in neurobiology where the latter play an important role. One example concerns the lateral diffusion of neurotransmitter receptors within the plasma membrane of a dendrite. A typical dendrite is studded with thousands of synaptic contacts, each of which corresponds to a local trapping region that binds receptors to scaffolding proteins, followed by internalization of the receptors via endocytosis \cite{Earnshaw06,Holcman06,Thoumine12}. The interior of a synapse thus acts as a partially-absorbing domain $\calM$ within the dendritic membrane $\Omega$. 
A related example is the passive or active intracellular transport of a vesicle (particle) along the axon or dendrite of a neuron, with absorption within a trapping region corresponding to the transfer of the vesicle to a synapse within the surface membrane of the neuron \cite{Bressloff21}. (In the latter case, the synapse could also be treated as a 2D absorbing surface in a 3D model of a neuron.) Both of these examples motivate extending
the analytical framework developed in this paper to investigate the competition for resources between multiple partially absorbing targets. 
As we have shown elsewhere \cite{Bressloff22a}, it is relatively straightforward to extend the generalized propagator BVP of section 2 to multiple domains $\calM_j$, $j=1,\ldots,N$, each with its own local absorption scheme. Now one has a set of occupation times $a_j$ and a corresponding set of Laplace variables $z_j$. One of the mathematical challenges is to develop efficient analytical or numerical schemes for inverting the solution with respect to the $N$ Laplace variables.

\vskip6pt

\enlargethispage{20pt}









\vskip2pc


 
\begin{thebibliography}{9}

\bibitem{Levy39} {L\`evy} P. 1939 {Sur certaines processus stochastiques homogenes.} {\em Compos. Math.} {\bf 7}, 283  

  \bibitem{McKean75} McKean HP. 1975 {Brownian local time.} {\em Adv. Math.} {\bf 15}, 91-111  
  
   \bibitem{Freidlin85} Freidlin M. 1985 {\em Functional Integration and Partial Differential Equations}
Annals of Mathematics Studies (Princeton University Press, Princeton
New Jersey)

\bibitem{Majumdar05}  Majumdar SN. 2005 Brownian functionals in physics and computer science. {\em Curr. Sci.} {\bf 89}, 2076  

\bibitem{Papanicolaou90} Papanicolaou VG. 1990 {The probabilistic solution of the third boundary
value problem for second order elliptic equations} {\em Probab. Th. Rel. Fields}
{\bf 87}, 27-77 

 \bibitem{Milshtein95} Milshtein GN. 1995 {The solving of boundary value problems by numerical
integration of stochastic equations.} {\em Math. Comp. Sim.} {\bf 38} 77-85

\bibitem{Singer08} Singer A, Schuss Z, Osipov A, Holcman D. 2008 {Partially reflected
diffusion.} {\em SIAM J. Appl. Math.} {\bf 68}, 844-868 

\bibitem{Bartholomew01}Bartholomew CH. 2001 Mechanisms of catalyst deactivation,
{\em Appl. Catal. A: Gen.} {\bf 212}, 17-60 


\bibitem{Filoche08}  Filoche M, Grebenkov DS, Andrade Jr JS, 
Sapoval B. 2008,Passivation of irregular surfaces accessed by
diffusion. {\em Proc. Natl. Acad. Sci.} {\bf 105}, 7636-7640


\bibitem{Grebenkov19b} Grebenkov DS. 2019  {Spectral theory of imperfect diffusion-controlled
reactions on heterogeneous catalytic surfaces}
{\em J. Chem. Phys.} {\bf 151}, 104108

\bibitem{Grebenkov20} Grebenkov DS. 2020  {Paradigm shift in diffusion-mediated surface phenomena.} {\em Phys. Rev. Lett.} {\bf 125}, 078102  


\bibitem{Grebenkov22} Grebenkov DS. 2022  {An encounter-based approach for restricted diffusion with a gradient drift.}  {\em J. Phys. A.} {\bf 55} 045203 

\bibitem{Bressloff22a} Bressloff PC. 2022  Diffusion-mediated absorption by partially reactive targets: Brownian functionals and generalized propagators. {\em J. Phys. A.} {\bf 55} 205001



\bibitem{Naim93} Ben-Naim E, Redner S, Weiss G. 1993 Partial absorption and virtual traps {\em J. Stat. Phys.}
{\bf 71}, 75



\bibitem{Earnshaw06} Bressloff PC,  Earnshaw BA. 2006 A biophysical model of {AMPA} receptor trafficking and its regulation during {LTP}/{LTD}.  {\em J. Neurosci.}{\bf 26}, 12362-12373 

\bibitem{Holcman06}
Holcman D, Triller A. 2006. Modeling synaptic dynamics driven by receptor lateral diffusion.
\newblock {\em Biophys. J.} \textbf{91}, 2405-2415 (2006).


\bibitem{Thoumine12}  Czondora K, Mondina M, Garciaa M, Heinec M,
Frischknechtc R, Choquet D, Sibaritaa JB, Thoumine OR. 2012 A unified quantitative model of {AMPA} receptor trafficking at synapses. {\em Proc. Nat. Acad. Sci. USA} {\bf 109} 3522-3527 

\bibitem{Bressloff21} Bressloff PC. 2021 Queuing model of axonal transport. {\em Brain Multiphysics} {\bf 2} 100042  




\end{thebibliography}
\end{document}